# Multi-Map Orbit Hopping Chaotic Stream Cipher

Xiaowen Zhang[1], Li Shu[2], Ke Tang[1]

**Abstract**
In this paper we propose a multi-map orbit hopping chaotic stream cipher that utilizes the idea of spread spectrum mechanism for secure digital communications and fundamental chaos characteristics of mixing, unpredictable, and extremely sensitive to initial conditions. The design, key and subkeys, and detail implementation of the system are addressed. A variable number of well studied chaotic maps form a map bank. And the key determines how the system hops between multiple orbits, and it also determines the number of maps, the number of orbits for each map, and the number of sample points for each orbits. A detailed example is provided.

## 1 Introduction

Chaotic cryptography has been attracting more and more attentions from the nonlinear system society since the essence of chaos dynamic system matches the very basic criteria of cryptography. The mixing of chaos matches the confusion of cryptosystem and sensitivity to initial conditions corresponds to the diffusion of cryptosystem. And the links to the conventional cryptography that has a wide range of appropriate design and analysis methods have been established [1] [2].

In the frequency hopping spread spectrum system, by following a specific hopping pattern the carrier of information "hops" from frequency to frequency over a wide band frequency spectrum. By doing so, noise-like signals are transmitted, and they are hard to be detected, intercepted, or demodulated.

Here we define a chaotic orbit [4]. Given $x_0 \in \boldsymbol{R}$ and a chaotic map $\boldsymbol{F}$, we define the orbit of $x_0$ under $\boldsymbol{F}$ to be the sequence of points $x_0$, $x_1 = F(x_0)$, $x_2 = F^2(x_0)$, $x_n = F^n(x_0)$, …. The point $x_0$ is called the seed of the orbit.

Rowlands [3] proposed an orbit hopping logistic map chaotic encryption scheme. In their algorithm, the key consists of three parameters: seed, settle, and offset. The seed is the initial value of the logistic map iteration sequence. The offset is the shift to the seed in order to generate a new orbit, i.e., the next orbit starts its map iteration from initial value of seed plus offset. The settle is some number of iteration times of a particular map.

Now let's explain a bit more about the settle and why we need it. In multiple orbit situations we want all orbits to be as dissimilar as possible. For one specific chaotic map, in order to have many orbits, a new orbit starts in another initial value deviated from its preceding one with a very small offset. A chaotic system is very sensitive to its initial value (it is well known butterfly effect), but it is a long term behavior. So the two orbits are almost overlapped each other in the beginning, which is bad thing for a cryptographic application. In order to avoid this situation from happening, before we use chaotic orbits / sample points for cryptographic purpose we will let adjacent orbits iterate some number of times (smaller offset may need more iterations) and make sure that they are different to each other.

---

[1] Graduate Center, the City University of New York, USA, E-mail: zhang_sean@yahoo.com
[2] College of Computer Science, Sichuan University, P.R. China

Inspired by frequency hopping, in our stream cipher the pseudo random numbers – the carriers in the sense of communication – are generated by hopping through many chaotic orbits by following a hopping pattern. Those orbits are generated from multiple chaotic systems / maps. The multi-map orbit hopping chaotic stream cipher is symmetric algorithm, it is fast, strong, computationally efficient, and applicable to most of applications, including wireless.

The organization of the paper is as follows. In section 2 we explain the design diagram of the cipher and chaotic map bank. Section 3 details the cipher implementation, key handling, subkey structure, and random number extraction method. In section 4 we give a concrete example of the cipher. Section 5 concludes the paper.

**2 Cipher Design**
Here we give the cipher design block diagram, and then verify that Chebyshev maps are chaotic maps.

2.1 Block Diagram

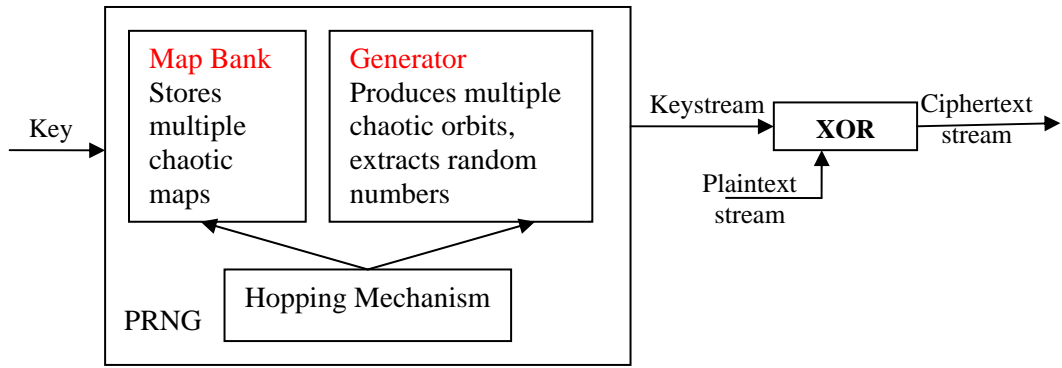

Fig 1. Block Diagram of Multi-Map Orbit Hopping Chaotic Stream Cipher

Given a key, the hopping mechanism performs a Key Handling process, then chooses $m$ maps $M_0, M_1 …, M_{m-1}$ from the chaotic map bank and sets the order of the chosen maps to hop. And for each individual chosen map, $s$ orbits $S_0, S_1, …, S_{s-1}$ are generated. And further, on each orbit, $n$ points $N_0, N_1, …, N_{n-1}$ are generated. Parameters $m$, $s$, $n$, and hopping pattern are determined by the key.

For a given chaotic map, the second orbit is generated by increasing the initial seed of the first orbit by an offset. The third orbit is generated by increasing the initial seed of the second orbit by the same offset, and so on so forth for other extra orbits needed.

2.2 Chaotic Maps Bank

For the pseudo random number generator of the stream cipher, we are using a bank of chaotic maps whose parameters are properly tuned to make sure that all maps lead to chaos.

In our current system, we are using a various number of logistic maps and Chebyshev maps, and both of them are well studied [4]. For logistic map
$$x_{n+1} = rx_n(1-x_n), \ x_n \in (0, \ 1)$$
where $0 \leq r \leq 4$. Most values of $r$ beyond 3.57 exhibit chaotic behavior (but there are still certain isolated values of $r$ that appear to show non-chaotic behavior, e.g. $r = 3.82$ ) [5].

For the Chebyshev map of degree $k$
$$x_{n+1} = \cos(2^k \cos^{-1} x_n), \quad -1 < x_n \leq 1, n = 1,2,3,...,$$
we do the transformation by
$$x = h(y) = 1 - 2y, \ y \in (0,1) \ ,$$
i.e.,
$$y = h^{-1}(x) = (1-x)/2 .$$
Substituting $x$ with the above expression of $y$ in the Chebyshev map, we have
$$1 - 2y_{n+1} = \cos(2^k \cos^{-1}(1-y_n)) .$$
Then we can have
$$y_{n+1} = (1 - \cos(2^k \cos^{-1}(1-y_n)))/2 .$$
When $r = 4$, logistic map $x_{n+1} = 4x_n(1-x_n)$ can be represented as the following equivalent format [6]
$$x_{n+1} = (1 - \cos(2\cos^{-1}(1-x_n)))/2 = L(x_n) .$$
So,
$$x_{n+k} = (1 - \cos(2^k \cos^{-1}(1-x_n)))/2 = L^k(x_n) .$$

Chebyshev map of degree k is the *k*th iteration of the logistic map. Therefore, Chebysheve maps possess all chaotic features that logistic maps have, they are ergodic, mixing, unpredictable, and extremely sensitive to initial conditions.

**3 Cipher Implementation and Key Issues**

3.1 Implementation Flowchart
Map Bank stores all possible chaotic maps that could be used for the system. Hopping Mechanism is in charge of picking up maps from Maps Bank, as well as key handlling and parameter controlling for Generator. The Generator consists of three loops. The outmost loop steps through all maps involved. For each map, several orbits are produced, further for each orbit several points are sampled to generate random numbers. The
number of maps, orbits, and points are determined by the key.

Notice that orbits for each map-loop will not be repeated. When coming back to the first
map-loop, orbits' offsets will be increased based on the offset the same map-loop left off
last time.

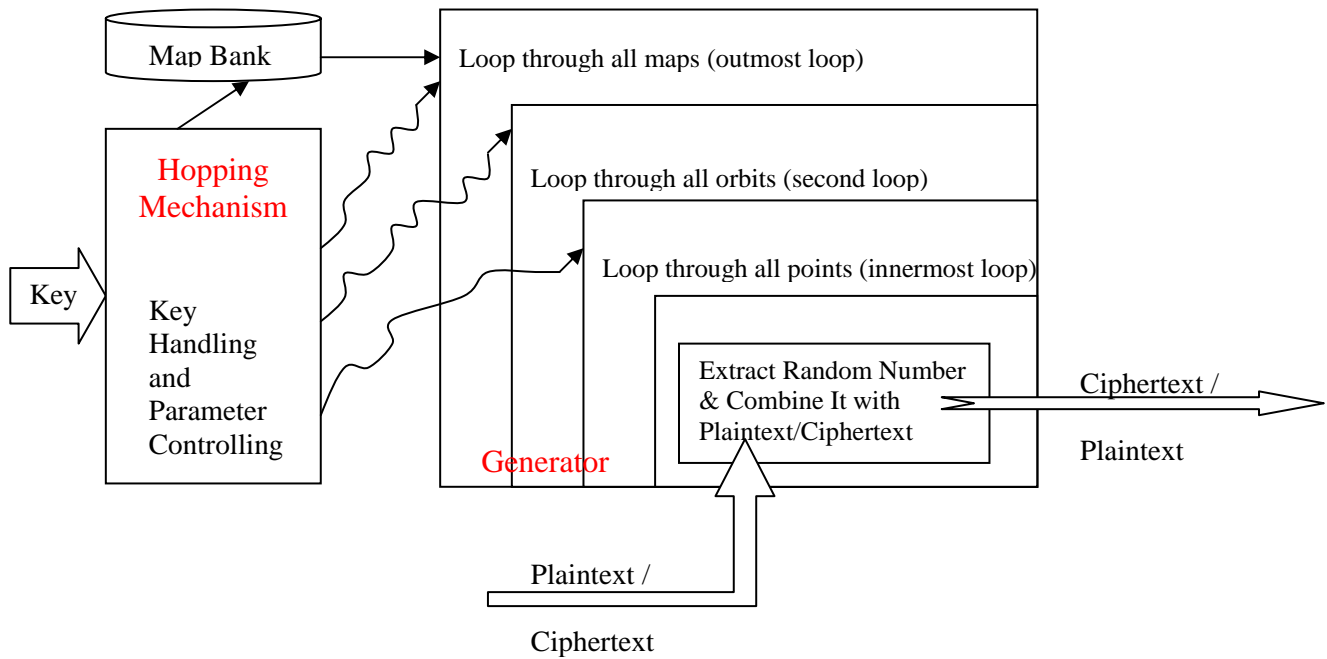

Fig 2. Detail Implementation Diagram

3.2 Key Handling
The key handling is the means by which the key bits are turned into subkeys that the
cipher can use. In this initial version we simply provide a much longer key and
sequentially split it into #maps subkeys. In the later version, key handling will use a
shorter key to generate all the subkeys needed for the system.

After a random-bit key is fed to the system, the key handling gets the #maps from the
first byte of the key. The rest of the key is split into #maps subkeys. And further all
control parameters are set up according to those subkeys. Assume that we can use slow
coin-flipping method to generate keys, it's one time work.

This is a variable-key-size stream cipher. The longer the key is, the stronger the cipher
will be.  The length of key is determined by number of maps involved in the cipher. For
the first version of the cipher, maps are picked up sequentially from the Map Bank
without further scrambling. Orbits are generated by increasing a unit Offset and sample
points are picked sequentially, too.

3.3 Subkey Structure and Key Length

Each map needs a subkey to control its behavior for generating suitable orbits. A subkey has the following structure.

| Seed | Offset | # Settles | # Orbits | # Samples |
|------|--------|-----------|----------|-----------|

Fig 3. Subkey Structure (# represents "number of")

The length of subkey is 56-bit. Here we explain the size of each component of subkey: Seed, Offset, #Settles, #Orbits, and #Samples.

- Seed: the initial value (state) of the chaotic map, represented in 6 hex-decimals, i.e. 24 bits. E.g., $(25fbd8)_h$ is the seed. Before being passed into map iteration, the hex will be converted into $(2489304)_d$ and then we add two zeros at left of decimal number, then add decimal point at the left of the zeros, makes it as 0.002489304.
- Offset: the shift value added to the Seed in order to generate the next orbit. It is represented in 4 hex-decimals, i.e. 16 bits. The Offset is far less than the Seed, at least 100 times smaller.
- #Settles: the number of iterations in settling period for an orbit before it can be used to generate sample points. It varies from 30 to 285. The variation is represented in 8 bits.
- # Orbits: the number of orbits generated during each use of a map. It is between 4 and 19, can be represented in 1 hex-decimal, i.e. 4 bits.
- # Samples: the number of sample points taken from an orbit, i.e. how long to stay on a particular orbit. It varies from 4 to 19 and is represented in 1 hex-decimal, i.e. 4 bits too.

The total key length is determined by the number of subkeys, because each map needs one subkey, i.e. the total key length is determined by the number of maps involved. In our current sittings the number of maps changes from 2 to 10, we introduce another 3 bits of uncertainty into the key. For instance, if we use 4 maps system, then the total key length will be (in bits) 3 + 4 x 56 = 227 bits. So the key space for the 4-map system is 2^227.

The random number extraction from a chaotic sample point is described in section 3.4. After the random number is generated, we combine it with plaintext / ciphertext to get corresponding ciphertext / plaintext. For the first version, we simply use exclusive OR (XOR) as the combination function (it could be replaced by other alternative.)

3.4 Chaotic Random Number Extraction

We extract the lowest digits from a specific chaotic orbit point $x_n$ by the following steps.

Step 1: Removes the decimal point from $x_n$ ($|x_n|$ is less than 1). E.g., 0.33461 → 33461, 0.9442345679457 → 9442345679457

Step 2: Makes the number 8 digits. If the original number is shorter than 8 digits, we add zeros at the end. If the original number is longer than 8 digits, we chop off the extra digits from the left side. E.g., 33461→33461000, 9442345679457 → 45679457.

Step 3: Gets the remainder from the 8 digits number by *mod* 256, it will be the generated pseudo random number. E.g., 33461000(*mod* 256) = 8, 45679457(*mod* 256) = 97.

The generated random numbers are in between 0 and 255. That's the total number of ASCII character set (including extended ones). That's for the simplicity of combining (XOR) random numbers with plaintext/ciphertext for encryption / decryption.

**4 An Example**

The system setup is as follows. A key is given to generate the chaotic pseudo random numbers. According to the key, the cipher uses eight maps. The key handling breaks the key into eight subkeys and sets up all the control parameters of the system.

Here is the key, it consists of 114 random numbers (0 to 15, i.e., 0 to F in hexadecimal). Notice that the first two numbers determine the #maps, and are generated separately. Here '1 11' is hex will be $(1B)_h$, i.e., 27. We calculate 2 + 27 (mod 7) = 8. So the #maps = 8. Each map needs a 56 bit subkey, so we need another 56 x 8 = 448 bits, i.e., 112 hex numbers for the rest of the key. Suppose those random numbers are generated by coin-flipping method.

"1 11   11 13 1 4 11 3 10 8 14 6 9 7 7   1 14 10 14 6 2 14 15 9 7 1 7 11 0   8 10 7 1 6 11 8 4 11 9 14 5 3 4   3 7 1 10 14 7 5 9 5 6 5 15 8 11 15 1 8 5 15 1 5 14 14 7 8 8 7 10   14 6 11 7 15 4 2 2 0 0 11 9 2 10   11 4 6 9 0 10 8 15 7 14 13 3 9 2   3 1 13 3 6 3 9 10 15 14 5 4 15 3"

Represented in hexadecimal format, the key will be: "1B BD144B3A8E6977 1EAE62EF9717B0 8A716B84B9E534 371AE759565F8B F185F15EE7887A E6B7F42200B92A B4690A8F7ED392 31D3639AFE54F3".

Here are the eight chaotic maps. The coefficients of those dynamical system difference equations are well tuned, and equations all lead to chaos.

0) Logistic map: $x_{n+1} = 3.901 x_n(1 - x_n)$,   $x_n \in (0,\ 1)$.

1) Logistic map: $x_{n+1} = 3.931 x_n(1 - x_n)$,   $x_n \in (0,\ 1)$.

2) Logistic map: $x_{n+1} = 3.963 x_n(1 - x_n)$,   $x_n \in (0,\ 1)$.

3) Logistic map: $x_{n+1} = 4 x_n(1 - x_n)$,   $x_n \in (0,\ 1)$.

4) Chebyshev map of degree 3: $x_{n+1} = \cos(2^3 \cos^{-1} x_n)$,   $x_n \in (-1,\ 1)$.

5) Chebyshev map of degree 4: $x_{n+1} = \cos(2^4 \cos^{-1} x_n)$,   $x_n \in (-1,\ 1)$.

6) Chebyshev map of degree 5: $x_{n+1} = \cos(2^5 \cos^{-1} x_n)$,   $x_n \in (-1,\ 1)$.

7) Chebyshev map of degree 6: $x_{n+1} = \cos(2^6 \cos^{-1} x_n)$, $\quad x_n \in (-1, 1)$.

According to the key, there are eight maps involved. Each of them has its own seed, offset, #orbits, #samples. The details are in the following table.

|        | Seed         | Offset       | #Settles | #Orbits | #Samples |
|--------|--------------|--------------|----------|---------|----------|
| Map #0 | 0.0012391499 | 0.00001499   | 135      | 11      | 11       |
| Map #1 | 0.002010722  | 0.000061335  | 53       | 15      | 4        |
| Map #2 | 0.009073003  | 0.000033977  | 259      | 7       | 8        |
| Map #3 | 0.003611367  | 0.00002287   | 125      | 12      | 15       |
| Map #4 | 0.0015828465 | 0.000024295  | 166      | 11      | 14       |
| Map #5 | 0.0015120372 | 0.00008704   | 30       | 4       | 4        |
| Map #6 | 0.001182337  | 0.000036734  | 241      | 13      | 6        |
| Map #7 | 0.003265379  | 0.000039678  | 114      | 19      | 7        |

There are 45 orbits come from four logistic maps (Map #0 ~ #3), and 47 orbits from the four Chebyshev maps (Map #4 ~ #7). We use different colors to distinguish orbits from different maps, Red for Map #0 / #4, Green for Map #1 / #5, Blue for Map #2 / # 6, and Black for Map #3 / #7. After their settle periods, the first four sample points of all those orbits are depicted in Figure 4. As shown here, those orbits are sufficiently different from each other. And furthermore we use our chaotic random number extraction method described in the previous section to obtain random numbers from those sample points.

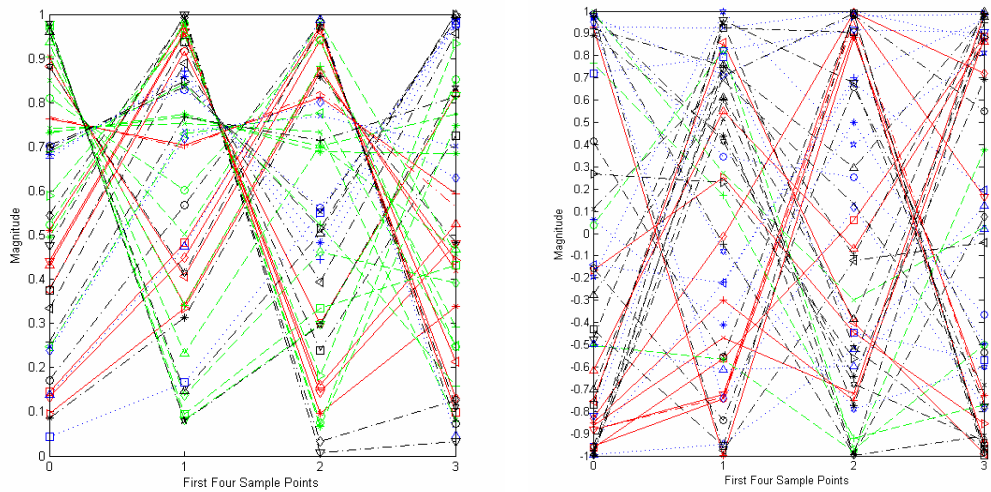

(a) Logistic Maps  (b) Chebyshev Maps
Fig. 4. 92 chaotic orbits of the first four sample points after their settle periods

Figure 5 and 6 are histograms to show frequencies of ASCII characters in plaintext (a) and ciphertext (b) for Mark Twain's novel "Huckleberry Finn" [7]. From figure 5(a) and 6(a), we see clustering around letters and peaking for space. In figure 5(b) and 6(b) we see pretty fairly even distribution, i.e., in the ciphertext novel each of 256 ASCII characters occurs with nearly equal likelihood.

The length of the novel in pure text format is 597,299 characters / bytes. The histograms of the plaintext and its corresponding ciphertext of the novel are as follows.

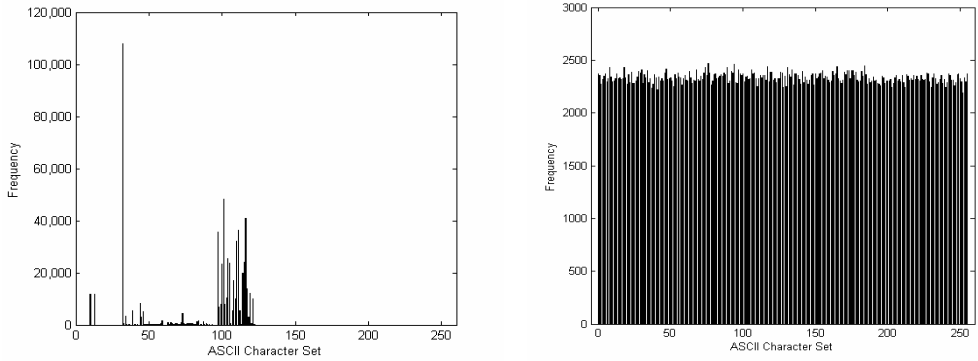

     (a) Novel in Plaintext         (b) Novel in Ciphertext
 Fig. 5. Histograms for Character Frequency of "Huckleberry Finn" in Pure Text Format

In MS Word format, the size of the novel is 1,404,928 characters / bytes. The histograms of the plaintext and its corresponding ciphertext encrypted by using the same key are provided here for comparison.

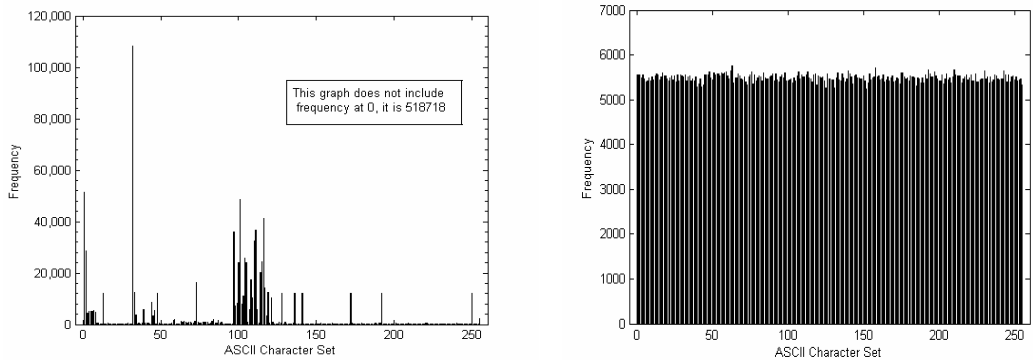

     (a) Novel in Plaintext         (b) Novel in Ciphertext
 Fig. 6. Histograms for Character Frequency of "Huckleberry Finn" in MS Word Format

**5 Conclusions**
In addition to chaotic features of mixing, unpredictable, and extreme sensitive to initial seeds, through multiple chaotic maps / orbits hopping mechanism, we spread out the pseudo random number base to a wide flat spread spectrum in terms of time and space. It is similar to say that our pseudo random numbers are out of the white noise. Therefore the distribution of those numbers is quite even, flattened, and further results in no clustering in the ciphertext as shown in figure 5 and 6.

The variable key length stream cipher that we have designed is a symmetric cryptosystem that makes use of chaotic system parameters as secret key: number of maps, seeds, offsets, settles, orbits, and sample points. Chaotic maps we chosen are computationally economic and fast, the current cryptosystem is easily implemented in software. We suggest that this cipher will be suitable for applications like wireless, cable, and optical fiber communications.

Final remark: the stream cipher will be tested against DIEHARD (a battery of tests of randomness) [8] and NIST statistical test suite [9]. Also cryptanalysis and security of the cipher remain to be conducted.

**Acknowledgements**
The authors would like to thank Prof. Michael Anshel for his encouragement and helpful comments. We also want to say special thanks to Prof. Kent D. Boklan for his review and corrections to the manuscript. Our thanks also give to Dr. Xiaowei Xu for valuable discussions.